\documentclass[%
 reprint,
superscriptaddress,
 amsmath,amssymb,
 aps,
prc,
]{revtex4-2}
\usepackage{todonotes,pifont}
\usepackage{graphicx}
\usepackage{dcolumn}
\usepackage{bm}
 \usepackage[%
pdftex,%
plainpages=false,%
colorlinks,%
linkcolor=blue,%
citecolor=blue,%
urlcolor=blue,%
linktocpage=true,%
breaklinks,%
pdfcreator=pdflatex,
pdfstartview=FitH,%
bookmarksopen=true
]{hyperref}

\begin{document}

\preprint{APS/123-QED}

\title{Properties of states near $E_x=6$ MeV in $^{18}$Ne through $^{17}$F$+p$ scattering}

\author{B. Sudarsan}
\email{sudarsanbalakrishnan01@gmail.com}
\affiliation{Department of Physics \& Astronomy, Louisiana State University, Baton Rouge, LA 70808, USA}
\affiliation{Department of Physics and Astronomy, Rutgers University, New Brunswick, New Jersey 08903, USA}

 \author{L. E. Linhardt}

\author{J. C. Blackmon}

\email{blackmon@lsu.edu}

\author{C. M. Deibel} 
 
\author{H. E. Gardiner}

\author{K. T. Macon} 

\author{B. C. Rasco}

\author{M. Matoš}

\affiliation{Department of Physics \& Astronomy, Louisiana State University, Baton Rouge, LA 70808, USA}

\author{D. Santiago-Gonzalez}
\affiliation{Physics Division, Argonne National Laboratory, Lemont, IL 60439, USA}
\affiliation{Physics Department, Florida State University, Tallahassee,
  FL 32306, USA}

\author{L. T. Baby}
\author{I. Wiedenh\"over}

\affiliation{Physics Department, Florida State University, Tallahassee,
  FL 32306, USA}

\author{E. Koshchiy}
\affiliation{Cyclotron Institute, Texas A \& M University, College Station, TX 77843, USA}

\author{G. Rogachev}
\affiliation{Department of Physics \& Astronomy, Texas A \& M University, College Station, TX 77843, USA}
\affiliation{Cyclotron Institute, Texas A \& M University, College Station, TX 77843, USA}
\affiliation{Nuclear Solutions Institute, College Station, TX 77843, USA}

\author{D. W. Bardayan}
\affiliation{Department of Physics and Astronomy, University of Notre Dame, Notre Dame, IN 46556, USA}
\newcommand{\comment}[1]{}
\date{\today}

\begin{abstract}
\begin{description}
\item[Background]The rate of energy production in the hot-CNO cycle and breakout to the rapid-proton capture process in Type I X-ray bursts is strongly related to the $^{14}$O($\alpha,p$)$^{17}$F reaction rate. The properties of states in $^{18}$Ne near $E_x=6.1-6.3$~MeV are important for understanding this reaction rate  \cite{WiescherAstro99}. 

\item[Experiment]The RESOLUT radioactive-ion beam facility at Florida State University \cite{wiedenhover2014} was used to study $^{18}$Ne resonances around this energy region using $^{17}$F(p,p)$^{17}$F elastic scattering on a polypropylene target under inverse kinematics. Scattered protons were detected in a silicon-strip detector array while recoiling $^{17}$F ions were detected in coincidence in a gas ionization detector. 

\item[Analysis]An $R$-matrix analysis of measured cross sections was conducted along with a reanalysis of data from previous measurements \cite{Hu14,Bla03,Bar12}. 

\item[Results]All the data analyzed are well described by a consistent set of parameters with  with a $1^-$ assignment for a state at 6.14(1) MeV. A second comparable solution is also found with a $3^-$ assignment for the 6.14(1) MeV state. The rate of the $^{14}$O($\alpha$,p)$^{17}$F reaction that is determined from the two solutions differs by up to an order of magnitude. 
\end{description}
\end{abstract}

\maketitle

\section{Introduction}
Type-I X-ray bursts (XRBs) are recurrent, explosive stellar phenomena occurring in interacting binary systems where matter accretes onto a neutron star from its companion.  Nuclear reactions in the accreted material initially proceed via the hot-CNO cycle, which is limited by the long $\beta$-decay lifetimes of $^{14}$O and $^{15}$O that are not burned by proton-induced reactions.
As temperatures increase, the $^{14}$O($\alpha,p$)$^{17}$F reaction can bypass the slow decay of $^{14}$O, leading to an increase in the rate of energy production and accelerating the thermonuclear runaway. Breakout from the hot-CNO cycle at higher temperatures can lead to the nucleosynthesis of isotopes up to the proton drip line and mass number $A \sim 100$. A recent computational study using a self-consistent, multi-zone XRB model found the $^{14}$O($\alpha,p$)$^{17}$F reaction as being one of the reactions that most influence the predicted light curves and burst ash composition \cite{Cyburt2016}.

The $^{14}$O($\alpha,p$)$^{17}$F reaction proceeds predominantly via natural-parity resonances close to the $\alpha$-threshold in $^{18}$Ne. It has long been believed that two of the most important resonances likely arise from states that are mirrors to the 6.20 MeV ($1^-)$ and 6.40 MeV ($3^-$) levels in $^{18}$O \cite{Wiescher87}. 
A variety of measurements have determined properties of states in $^{18}$Ne including three energy levels at $E_x=$~6.14, 6.29 and 6.35 MeV \cite{Hah96,Gom01,Bla03,Not04,He09,Alm12,Hu14,Thom18}. The level-density in this region is low, and it is clear that two of these levels are the $1^-$ and $3^-$ states of interest, while the other is the mirror to the $2^-$ state at 6.35 MeV in $^{18}$O that does not contribute to the $^{14}$O($\alpha,p$)$^{17}$F reaction rate due to its unnatural parity. Reported properties of states in $^{18}$Ne in the region are summarized in Table \ref{resonancetable}. 

Conclusive spin-parity identification of these three levels and a determination of the partial widths are important to reduce the uncertainties in the $^{14}$O($\alpha,p$)$^{17}$F reaction rate.  There are many open particle-decay channels for states in the region of interest: proton emission to the ground state of $^{17}$F, proton emission to the first-excited state of $^{17}$F, direct two-proton emission to $^{16}$O, and $\alpha$ emission. The alpha-partial widths ($\Gamma_{\alpha}$) most directly impact the reaction rate, but the other partial widths are also important for interpreting measurements of the time-inverse $^{17}$F($p,\alpha$)$^{14}$O reaction that can determine resonance strengths of interest.

Thomas-Ehrman level-shift calculations favor the lowest-energy states of these three levels in each nucleus as mirrors, pairing the 6.14-MeV state in $^{18}$Ne with the 6.20-MeV ($1^-$) state in $^{18}$O. Angular distributions from the $^{16}$O($^3$He,$n$)$^{18}$Ne reaction are more consistent with $\ell \le 2$ for the 6.14-MeV state \cite{Hah96}. Reaction selectivity arguments also favor the highest-energy state having unnatural spin-parity \cite{Hah96,Gom01,He10,Thom18}. These observations have led to the tentative assignment of 6.14 ($1^-$), 6.29 ($3^-$) and 6.35 ($2^-$) MeV in $^{18}$Ne, which we refer to as the conventional spin ordering. However, there is also evidence to the contrary, most notably angular distributions from the $^{20}$Ne($p,t$)$^{18}$Ne reaction preferring a $3^-$ assignment for the 6.14-MeV state \cite{Thom18}.

A recent study of the $^{17}$F$ + $p elastic scattering cross section reported three resonances in the region, determined their properties based on an $R$-matrix analysis, and also argued for a preference for the conventional spin ordering of states \cite{Hu14}. Definitive $J^{\pi}$ assignments of $3^-$ and $1^-$ from elastic scattering are challenging given the same orbital-angular-momentum transfer for each of these levels. In this paper we report on a new measurement with improved energy resolution of $^{17}$F$+ $p elastic scattering using a solid polypropylene (CH$_2$)$_n$ target. We also present results from a comprehensive $R$-matrix analysis of cross section data from the current work and data from the previous studies of Ref. \cite{Hu14, Bar12}. 

\begin{table*}[t]
\caption{\label{resonancetable}Summary of previously reported resonance energies and widths for $^{18}$Ne states with $E_x=6-7$ MeV. } 

\begin{ruledtabular}
	\begin{tabular}{c c c  c c  c c  c c  c c  c c}
 & \multicolumn{2}{c}{Hahn \textit{et al} \cite{Hah96}} &  \multicolumn{2}{c}{Del Campo \textit{et al }\cite{Gom01}} &  \multicolumn{2}{c}{Park \textit{et al} \cite{Par99}} & \multicolumn{2}{c}{Harss \textit{et al} \cite{Hars02}} &  \multicolumn{2}{c}{Bardayan \textit{et al} \cite{Bar12}}&  \multicolumn{2}{c}{Hu \textit{et al} \cite{Hu14}} \\
$J^{\pi}$& $E_x$  & $\Gamma_p$  &  $E_x$  & $\Gamma_p$ &  $E_x$  & $\Gamma_p$ &  $E_x$  & $\Gamma_p$  &  $E_x$  & $\Gamma_p$ &  $E_x$  & $\Gamma_p$  \\
&  MeV  & keV  &  MeV  & keV  &  MeV  & keV &  MeV  & keV &  MeV  & keV & MeV  & keV \\
 \hline
$(1^-)$ &6.150(10) & $\leq$40 & 6.143(1) &50(5)& && && 6.134(1)& 54(2)& 6.15(3)&50(15)\\
$(3^-)$ &6.293(10) & $\leq$20 & && 6.305(4) & 8(7) &  && & & 6.28(3)&20(15)\\
$(2^-)$& 6.345(10) & 45(10) &6.343(1) & 50& 6.358(5) &18(9)& && & & 6.35(3)&10(5)\\ 
$0^-$/$0^+$ & & && & && && & & 6.85(11)&50(30)\\
$4^+$  &7.07(10)& 200(40) & && && 7.05(10)&90(40)& & & 7.05(3)&95(20)\\ 

\end{tabular} 
\end{ruledtabular}
\end{table*}

\section{Experiment}
The $^{17}$F$+$p elastic scattering cross section was measured at the John D. Fox Superconducting Accelerator Laboratory at Florida State University (FSU). A 54.7-MeV beam of $^{17}$F was produced in-flight using the $^{16}$O($d,n$)$^{17}$F reaction with a 79-MeV beam of $^{16}$O bombarding a cryogenically-cooled deuterium gas cell with 2-$\mu$m-thick havar windows. The primary beam and heavy-recoil products were collected downstream with a large-bore solenoid magnet and separated to a purity of about 70\% $^{17}$F by the RESOLUT system and  intensity of $10^5$ s$^{-1}$.

The secondary $^{17}$F radioactive beam bombarded a 2.05-mg/cm$^2$ polypropylene target, and scattered protons were detected by a telescope of double-sided silicon detectors of Design S2 fabricated by Micron Semiconductor. The first $\Delta$E layer was 65-$\mu m$ thick, while the residual particle energy was measured by a 1-mm-thick backing detector. The detectors were positioned 6.6 cm downstream of the target, covering $\theta_{lab}=8^{\circ}-24^{\circ}$.

\begin{figure}[htp]
 	\includegraphics[scale=0.4]{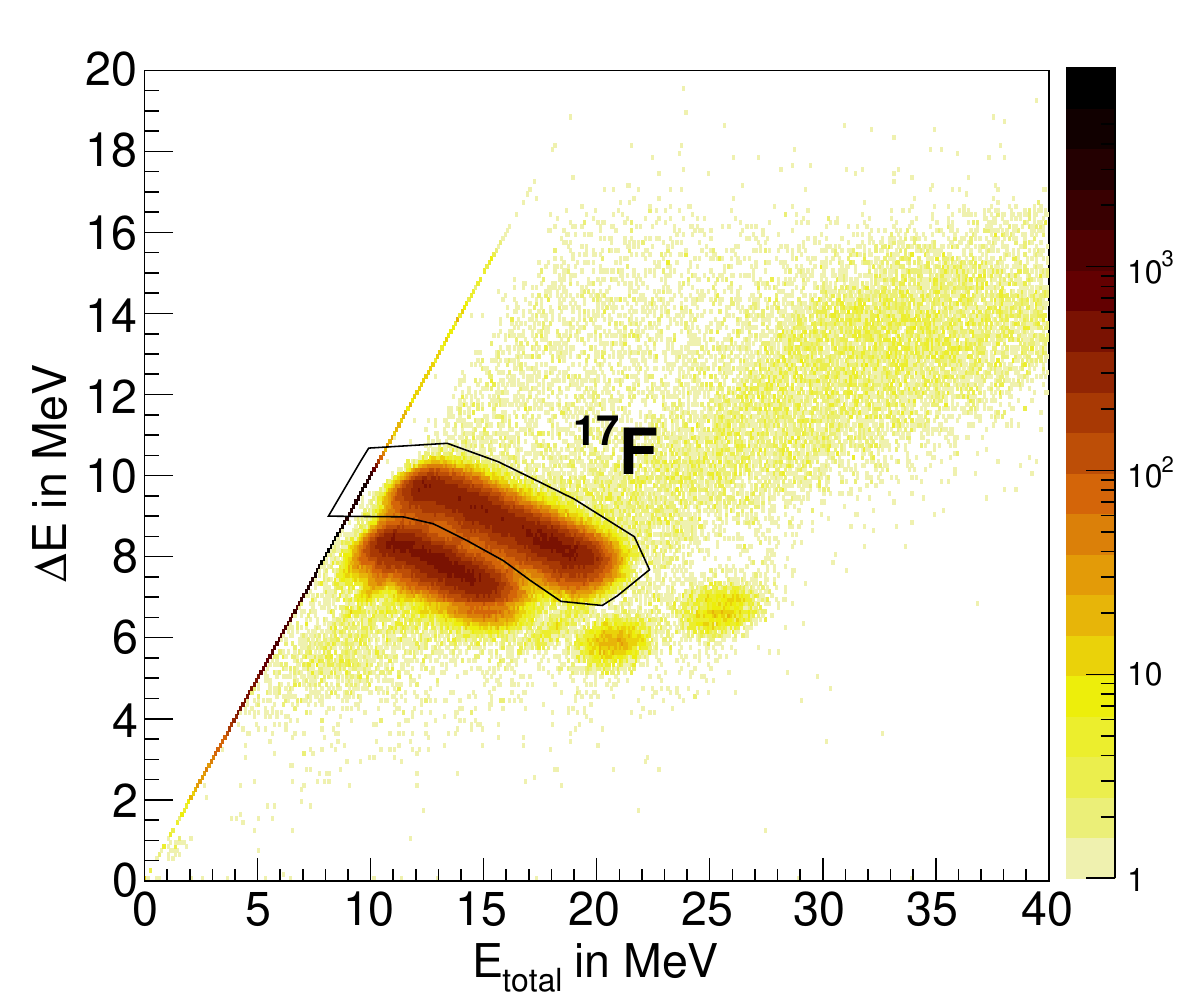}	
  \caption{\label{fig:ExpFigure} Particle identification plot (energy lost in the first 12 cm of gas, $\Delta E$, versus total energy lost $E_{\rm total}$) from the gas ionization chamber for events in time coincidence with the silicon-strip detector.}
\end{figure}

Recoiling heavy ions were detected in coincidence by a gas ionization chamber located 25.7 cm downstream from the target \cite{Lai18}. The ion chamber has a 6.4-cm-diameter window made of 7-$\mu$m-thick kapton centered on the beam axis. A 1.0-cm-diameter solid aluminum disk suspended in the middle of the window covered $\theta_{\rm lab}<1.1^{\circ}$, preferentially blocking the unscattered beam and minimizing pile-up of events in the ionization chamber. The ion chamber was filled with isobutane gas at 15 Torr, and recoiling heavy ions were identified by their energy loss collected in the first 12-cm depth of gas, $\Delta E$, compared to the total energy, $E_{\rm total}$.
The events of interest were selected in part by gating on the $^{17}$F band in the $\Delta E$-$E$ particle identification plot from the gas ionization chamber as shown in Fig. \ref{fig:ExpFigure}. A time coincidence was required with events in the silicon telescope that registered energy signals greater than 4~MeV in the 1-mm-thick detector. This combined set of conditions unambiguously selected $^{17}$F+p scattering events. 

 Particle energies in the silicon-strip detector were calibrated using a $^{228}$Th alpha source. Some of the highest-energy protons did not register a reliable energy signal above threshold in the 65-$\mu$m-thick detector. Therefore, the center-of-mass energy for each event, $E_{\rm cm}$, was determined based upon the energy and angle of the proton detected in the 1-mm-thick detector, while correcting for the energy lost by the proton in the 65-$\mu$m-thick detector using calculated values from LISE++\cite{Lise++}. Thus, we treated the 65-$\mu $m-thick detector as an inert absorber in reconstructing the $E_{\rm cm}$. The calculated energy loss for protons in the $\Delta E$ detector was tested by comparing to the majority of protons that did provide a reliable signal in the $\Delta E$ detector and was found to be in good agreement. As a further check on the $E_{\rm cm}$ determination, the highest-energy protons detected result from scattering in the upstream edge of the target, and we used these protons to reconstruct the incident beam energy finding $55.1 \pm 0.5$~MeV in good agreement with the beam energy (54.7 MeV) expected from the RESOLUT magnetic field settings.

\begin{figure}[htp]
	\includegraphics[width=\columnwidth]{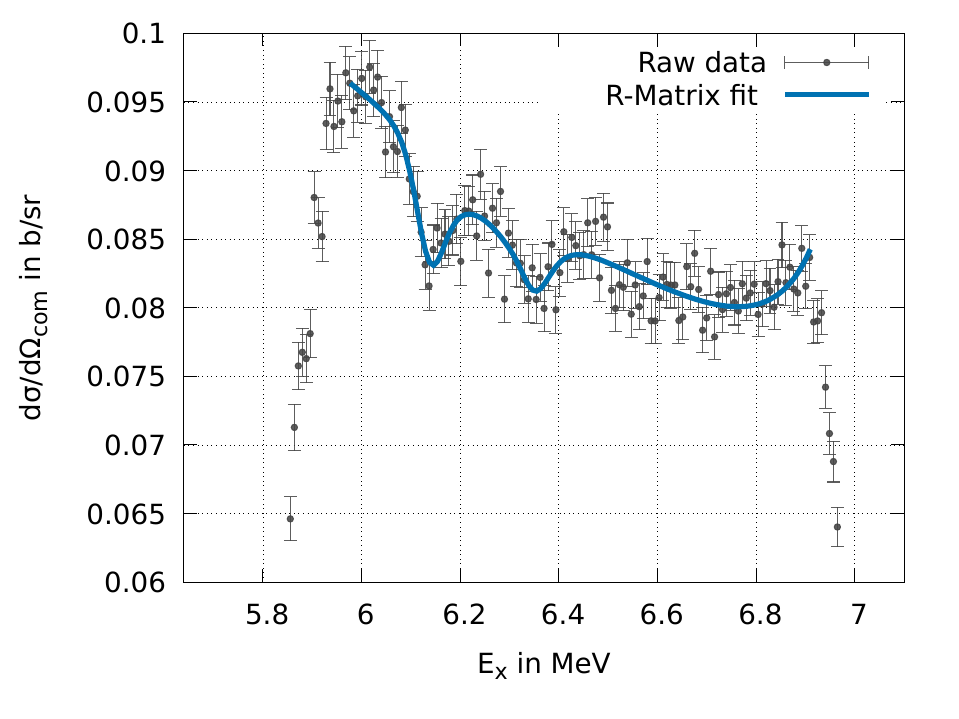}
	\caption{The individual data points are differential cross-sections in b/sr in the COM frame for $^{17}$F + p elastic scattering measured at $\theta_{cm} = $ 146.9$^{\circ}$. Also shown is an $R$-matrix fit with a $1^-$ assignment for the state at 6.14-MeV and a $3^-$ assignment for the state at 6.29 MeV. The fit parameters are provided in Table \ref{tab:table3}.
	}
 \label{fig:xsection-146}
\end{figure}

The yield of scattered protons versus $E_{\rm cm}$ was converted into a differential cross section normalizing to $^{12}$C($^{17}$F,$^{17}$F)$^{12}$C elastic scattering detected by the gas ionization chamber assuming Rutherford scattering with the solid angle calculated from geometry. The $^{17}$F + p excitation function converted to excitation energy ($E_x$) in $^{18}$Ne is shown in Fig. \ref{fig:xsection-146}, which spans
$E_x \approx 5.9-6.9$~MeV. While the absolute overall normalization has modest uncertainty (about 30\%) limited by the beam current normalization, the relative cross section as a function of energy is quite precise, and two resonant features are observed in the excitation function at excitation energies of about 6.15 and 6.35 MeV.  The detected protons have an overall energy resolution of 42-keV FWHM in the center-of-mass frame, resulting from the intrinsic energy resolution of the Si telescopes, the effect of proton straggling in the $\Delta$E layer, and the contribution due to the detector's angular resolution. 

\section{$R$-matrix Analysis}
We performed an $R$-matrix analysis of measured $^{17}$F+p elastic-scattering cross sections to determine properties of states in $^{18}$Ne using the multichannel $R$-matrix software AZURE2 \cite{Azuma10}. A suitably large channel radius of $r = 4.864$~fm ($r_0 = 1.36$~fm) was chosen to be outside the interior interaction region due to the loosely-bound nature of the system. Data were fit with $R$-matrix calculations at a single angle ($\theta_{cm} = $ 146.9$^{\circ}$) near the center of the angular range covered. The calculated cross-section profile was smeared by a Gaussian kernel with width matching the experimental energy resolution.

The properties of states with $E_x$ $<$ 6 MeV (below the energy range of interest) are generally well known, and all of these states were included in the analysis with their resonance parameters fixed to previously reported values \cite{Hah96,Par99,Jun10,Bar99,Par05}.
The important states in this region are at 4.52 and 5.11 MeV, corresponding to $s$-wave resonances. We also include states that would form $p$-wave and $d$-wave resonances in the calculation, though they have a negligible influence in the region of interest.  For proton partial widths of these resonances not previously measured, we fixed their values at estimated upper limits following the formalism discussed in Ref. \cite{WignerLimitPaper}. 
To reduce the number of parameters, we include only channels that correspond to the lowest possible orbital angular momentum for each state, which is likely dominant due to the reduced penetrability for higher orbital angular momenta.

There is more uncertainty about the properties of resonances with $E_x>$ 7 MeV (higher in energy than the region of interest). Some of these states are broad and influence the overall magnitude and general trend with energy of the scattering cross section in the region of interest. We include a previously observed resonance at 7.37 MeV \cite{Tilley95,Hah96,Hars02,Fortune03,Not04} with $J^\pi=2^+$ and a width of 40 keV.   
Two other states at 7.6 and 7.7 MeV are also included with their widths fixed to previously reported values \cite{Hars02}. 
We also included other broad hypothetical states fixed at higher energies to provide a smoothly varying background contribution and mimic the tail of the continuum. The partial widths for background state(s) and the overall normalization of the data were varied as free parameters. 

An important resonance affecting the overall shape of the cross section in the region of interest results from a 4$^+$ state at about 7.07 MeV in $^{18}$Ne that was proposed \cite{Funck88,Sherr98,Fortune00,Fortune03} and verified \cite{Tilley95,Hah96,Hars02,Not04,He11,Hu14} in previous work. The previous $^{17}$F+p elastic scattering data of Ref. \cite{Hu14} covers a broader range of energies than in our study, including the region of the $4^+$ resonance. We reanalyzed the data from Ref. \cite{Hu14} to better constrain the influence of the 7.07-MeV state in the region of interest and determine if a consistent set of resonance parameters could describe both scattering measurements. 

\begin{figure}[htp]
	\includegraphics[width=\columnwidth]{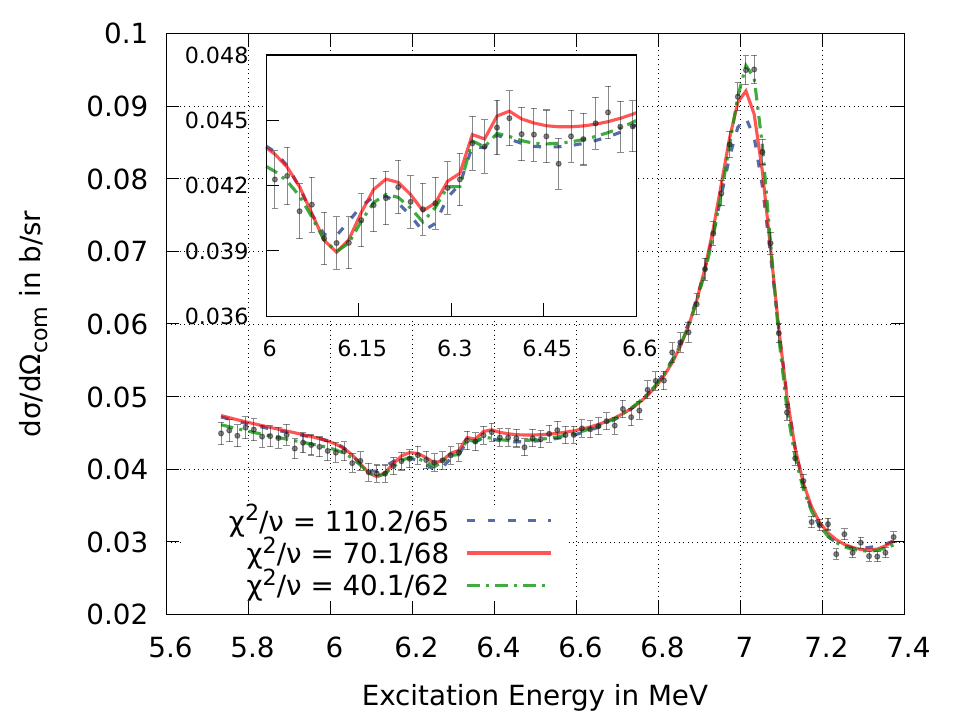}
	\caption{\label{fig:xsection-155}Differential cross-section in b/sr in the COM frame for $^{17}$F + p elastic scattering from Ref. \cite{Hu14} at a reported angle of $\theta_{cm} \approx 155^{\circ}\pm 18^{\circ}$. $R$-Matrix fits are shown as coloured lines here with the conventional spin ordering.  The first fit (blue, dashed) assumes $\theta_{cm}=$ 156.4$^{\circ}$ and uses three different background resonances with a fixed $E_x=8.9$ MeV, varied $\Gamma_p$, and $J^\pi=3^-, 4^+$ and $2^-$ respectively. The second fit (red) assumes $\theta_{cm}=$ 167.8$^{\circ}$ and uses a single background resonance with $J^\pi=3^-$ and same assumptions for the $E_x$, $\Gamma_p$ values as in the first fit. The third fit (green, dot-dashed) makes the same assumption as in the first, but fits the region around $E_x=7.05$ MeV with two $J^\pi=4^+$ resonances instead of one, both with freely varying parameters $E_x$ and $\Gamma_p$. } 
\end{figure}

An excitation function for this data \cite{Hu14} is shown in Fig. \ref{fig:xsection-155} at one angle denoted as $\theta_{lab} = 3^{\circ}$ and $\theta_{cm} \approx 155^{\circ}\pm18^{\circ}$ in Ref. \cite{Hu14}, along with 3 different fits. Our best fit using a single $4^+$ state near $E_x \approx 7$~MeV and calculated at $\theta_{cm} = 156.4^{\circ}$ (blue, dashed) does not fit the peak of the $4^+$ resonance well and results in $\chi^2/\nu=1.69$. Assumptions about the fit were adjusted to try to improve the quality of the fit, as illustrated in Fig.\ \ref{fig:xsection-155}. Shifting the angle of the calculation to $\theta_{cm} = 167.8^{\circ}$ results in a significantly better fit to the data with $\chi^2/\nu=1.03$. This is well within the range of angles covered in the measurement and may indicate that the relatively poor quality of the original fit results from the simplification of calculating at a single scattering angle rather than an angle-integrated cross section. Previous work has suggested that 2 unresolved states separated by about 70 keV could contribute around $E_x \approx 7$~MeV \cite{Hah96,Not04}, though this is not firmly established. We find even better fits ($\chi^2/\nu=0.65$) to the data of Ref. \cite{Hu14} when including two such resonances (green, dot-dashed), in contrast to the findings in Ref. \cite{Hu14}. It should be noted that this data includes systematic uncertainties, so $\chi^2/\nu<1$ is not unexpected. 

The best fit values for the $4^+$ state from an analysis of the data from the 2 angles of Ref. \cite{Hu14} is shown in Table II. We find consistent widths for the 4$^+$ state, but resonance energies that differ by about 30 keV. We adopt average values ($E_x=7.065$~MeV and $\Gamma_p=151$~keV) in our analysis of the FSU data set. We find a strong preference for population of the $4^+$ state through the $S_{\text{tot}}=2$ spin channel, which results in a significantly larger width than was reported in Ref. \cite{Hu14}, where a strong $S_{\text{tot}}=3$ spin channel was used. We also did not find evidence for a suggested $0^+/0^-$ state at E$_x=6.85$ MeV \cite{Hu14}, as its inclusion did not significantly improve the $\chi^2$ in our analysis in either data set.  For the resonance at 7.37 MeV with width fixed at 40 keV, we observed statistically significant improvements to the total $\chi^2$ when assigning $J^\pi=0^+$ instead of $2^+$, and this assignment has been used in all the results in Table II.

\begin{figure}[htp]
    \centering
	\includegraphics[width=\columnwidth]{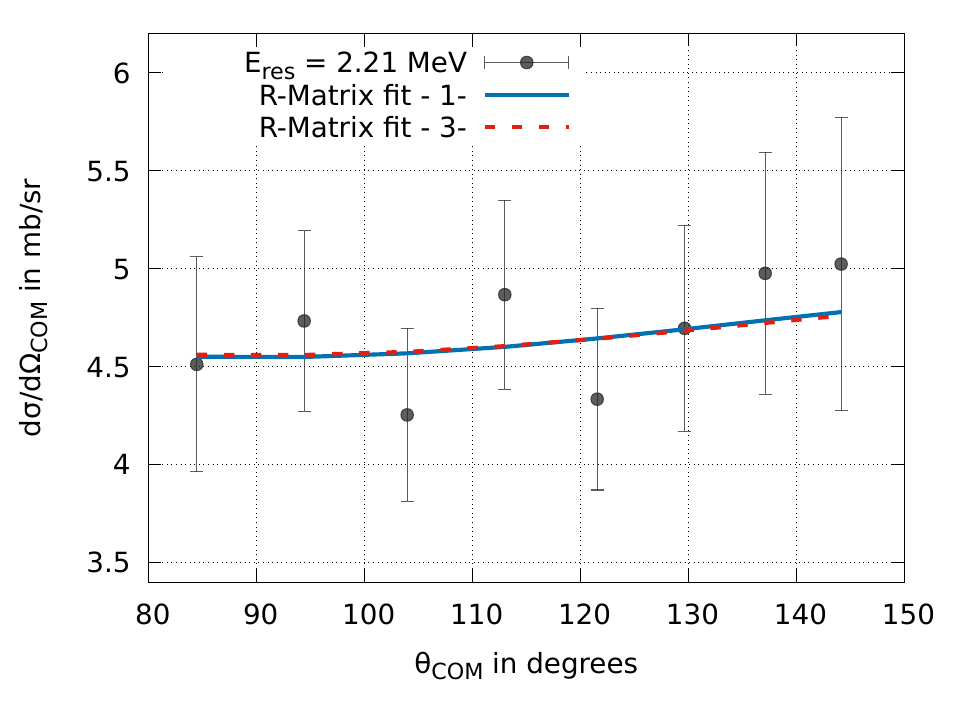}
	\caption{\label{fig:ORNLAngDist} Angular distribution of inelastic $^{17}$F(p,p$^\prime$)$^{17}$F$^\star$ data from reference \cite{Bla03,Bar12}, with $R$-Matrix fits using $J^{\pi}$ of  $1^-$ (blue, solid line) and $3^-$ (red, dashed line) for the state at $E_x=$6.14 MeV.}
\end{figure}

\begin{figure}[htp]
    \centering
    \includegraphics[width=\columnwidth]{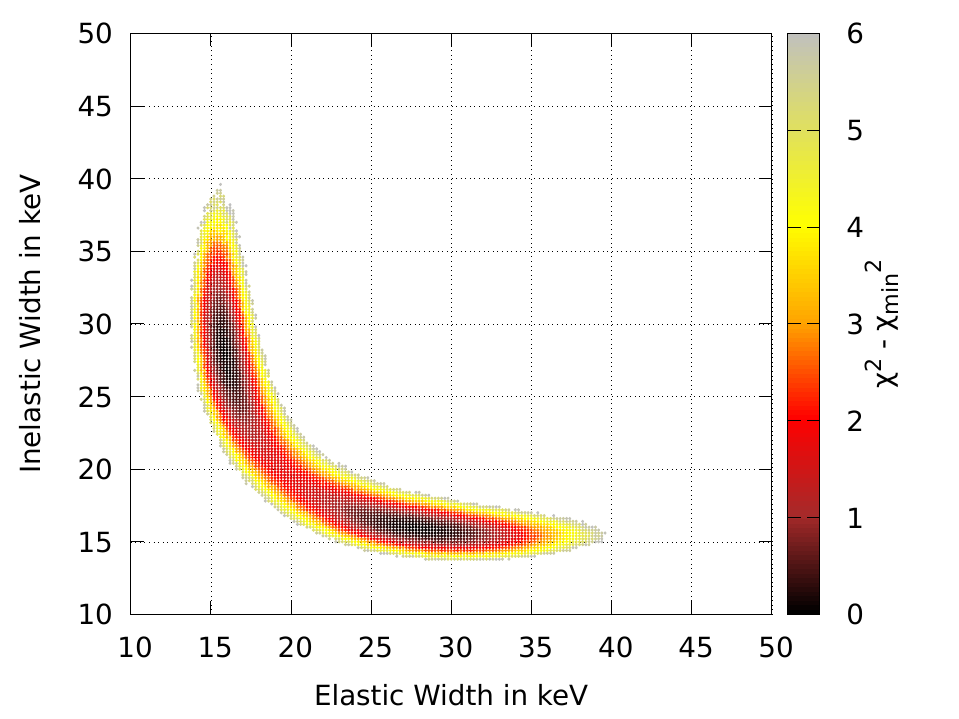}
	\caption{\label{fig:walkers} Surface plot showing the change in $\chi^2$ when fitting the ORNL dataset with the $(1^-,3^-)$ ordering, against the elastic and inelastic widths of the $1^-$ state.}
\end{figure}

\begin{table*}
\renewcommand*\arraystretch{1.3} 
 \caption{\label{tab:table3}Comparison of $R$-Matrix best-fit results for states in the region E$_x = 6.1-6.4$ MeV.
	Since the state with $J^{\pi}=4^+$ was not measured/constrained in this work, we do not report its average properties.}
	\begin{ruledtabular}
		\begin{tabular}{c | cccc | ccc | ccc | ccc | c}
    Dataset&$E_x$&J$^{\pi}$&$\Gamma_{\text{total}}$ & $\Gamma_{p^{\prime}}$&$E_x$&J$^{\pi}$&$\Gamma_p$&$ E_x$&J$^{\pi}$&$\Gamma_p$&$ E_x$&J$^{\pi}$&$\Gamma_p$&$\chi^2/\nu$\\
			&MeV&&keV& keV &MeV&&keV&MeV&&keV&MeV&&keV&\\ \hline
		
	           This work&6.134(8)&$1^-$&51$_{-12}^{+18}$&15\footnote{Fixed in $R$-matrix fit}&6.30(5)&$3^-$&$<$36&6.35$_{-2}^{+3}$&$2^-$&90$_{-36}^{+55}$\footnote{From fit without 3$^-$ included}&  7.12(2)&4$^+$&156\footnotemark[1]&0.99\\
			
	           \cite{Hu14} 3$^{\circ}$&6.121(10)&$1^-$&47$_{-9}^{+11}$&15\footnotemark[1]&6.296$_{-11}^{+3}$&$3^-$&11(4)&6.348$_{-5}^{+4}$&$2^-$&$<$21&7.043(2)&4$^+$&152$_{-5}^{+4}$&1.03\\
            
             \cite{Hu14} 10$^{\circ}$&6.17(2)&$1^-$&54$_{-13}^{+19}$&15\footnotemark[1]&-\footnote{No clear central value observed}&$3^-$&$<$8&6.356$_{-10}^{+15}$&$2^-$&13$_{-6}^{+25}$&7.0739$_{-21}^{+14}$&4$^+$&150(6)& 0.68\\			 

            ORNL\cite{Bla03,Bar12}&6.1321$_{-11}^{+10}$&$1^-$&44$_{-4}^{+5}$&25$_{-8}^{+7}$&&&&&&&&&&0.99\\
            
            Average&6.132(1)&$1^-$&46(4)&&6.297$_{-10}^{+3}$&3$^-$&11(4)&6.349$_{-4}^{+3}$&2$^-$&60$_{-22}^{+32}$&&&&\\
            \hline

            This work&6.145(6)&$3^-$&10$_{-2}^{+3}$&5\footnotemark[1]&6.342$_{-15}^{+10}$&$1^-$&15$_{-15}^{+11}$&6.406$_{-28}^{+13}$&$2^-$&16$_{-13}^{+12}$&  7.12\footnotemark[1]&4$^+$&152\footnotemark[1]&1.18\\

			\cite{Hu14} 3$^{\circ}$&6.136(4)&$3^-$&18$_{-4}^{+5}$&5\footnotemark[1]&6.253(12)&$1^-$&35$^{+11}_{-9}$&6.34(4)&$2^-$&$<$65&7.043(2)&4$^+$&151$_{-5}^{+4}$&1.06\\
			
			\cite{Hu14} 10$^{\circ}$&6.17$_{-3}^{+2}$&$3^-$&13$_{-4}^{+8}$&5\footnotemark[1]&6.247$_{-4}^{+6}$\footnote{Not included in average reported, due to mismatched spins}&$2^-$&5$_{-5}^{+31}$\footnotemark[4]&6.354$_{-15}^{+14}$\footnotemark[4]&$1^-$&54$_{-16}^{+23}$\footnotemark[4]&7.074(2)&4$^+$&151(6)&0.70\\
			
            ORNL\cite{Bla03,Bar12}&6.1415(6) &$3^-$& 21(2) & 4.7(2) &&&&&&&&&&1.00\\

            Average&6.142(1)&$3^-$&19(2)&&6.266(12)&1$^-$&27(5)&6.38$_{-4}^{3}$&2$^-$&16$_{-13}^{+12}$&&&&\\
        \end{tabular}
	\end{ruledtabular}
\end{table*}

\begin{figure*}[]
	\includegraphics[scale=1.0]{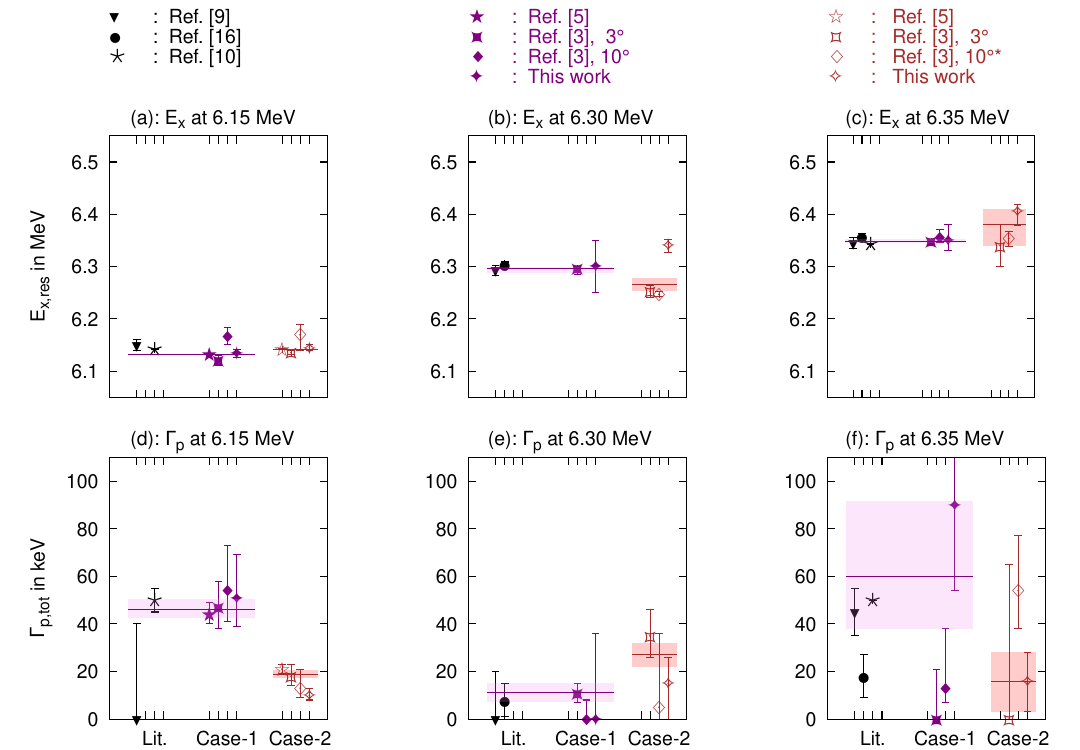}	
     \caption{\label{fig:summary}Summary of all assignments to the three resonances around $6 - 6.5$ MeV in $^{18}$Ne. Violet points (labeled `Case-1') denote the case of the 6.15-, 6.30-, and 6.35-MeV states assigned spin parities of $1^-$, $3^-$, and $2^-$, respectively, the red points (labeled `Case-2') denote the case of the 6.15-, 6.30-, and 6.35-MeV states assigned spin parities of $3^-$, $1^-$, and $2^-$, respectively, and the black points (labeled `Lit.') denote prior measurements, all of which used the $(1^-,3^-,2^-)$ order. The bands denote the 68\% confidence interval combining all the measurements with a matching color according to the linear variance prescription by Barlow \cite{barlow2004asymmetric}, unless indicated otherwise.  *This data was not included in the $E_x$ and $\Gamma_p$ averages for the 6.30-, and 6.35-MeV states shown in the red band as the spin ordering for this case was ($3^-,2^-,1^-$).}
\end{figure*}

\section{Analysis for states with $E_x=6-7$ MeV}

Within the region of interest, three resonances have been widely reported at 6.14, 6.29, and 6.35 MeV, with their excitation energies determined to an uncertainty of about 0.01 MeV through transfer reactions. These studies also indicate that the width of the states is smaller than the instrumental resolution of about 50 keV. We included three resonances in this energy region in our analysis, with spin-parities of $1^-$, $2^-$, and $3^-$ as indicated by mirror symmetry. 

A previous measurement of $^{17}$F(p,p$^\prime$)$^{17}$F$^\star$ inelastic scattering at Oak Ridge National Laboratory (ORNL) showed the state near E$_x$ = 6.14 MeV to have a significant partial proton width to the 495-keV first-excited state in $^{17}$F \cite{Bla03,Bar12}.
Since the elastic scattering data does not provide a good constraint on the width to the first-excited state in $^{17}$F $(\Gamma_{p^{\prime}})$, we also included data from the ORNL measurement in our $R$-matrix analysis. The ORNL data near the peak of the resonance were divided into angular bins, and we fit the differential cross sections instead of the total cross section as was done in Ref. \cite{Bar12}. However, the differential cross sections are nearly isotropic (see Fig. \ref{fig:ORNLAngDist}), and the results from our $R$-matrix fit (given in Table \ref{tab:table3}) are consistent with those of Ref. \cite{Bar12}.  

The change in $\chi^2$ is plotted in Fig. \ref{fig:walkers} for the $J^{\pi}= 1^-$ assignment of the 6.14-MeV state as a function of the partial elastic and inelastic widths.
This shows there is no preference in the fits to the inelastic scattering cross section as to whether the elastic or inelastic width is larger. As the branching ratio measurements of Ref. \cite{Alm12} rule out a large inelastic scattering width, a larger ground-state proton partial width is adopted as would also be expected from penetrability arguments. 

We analyzed the elastic-scattering excitation function from Fig. \ref{fig:xsection-146} and the two angles reported in Ref. \cite{Hu14} independently, but with a consistent approach. The inelastic width $(\Gamma_{p^{\prime}})$ for the state near 6.14-MeV was included as a fixed parameter with its value taken from the analysis of the ORNL data described above. The energies and other partial widths for the $1^-$, $2^-$, and $3^-$ resonances in the region of interest were varied including all possible channel-spin contributions corresponding to the lowest possible orbital angular momentum for each channel. We also varied the properties of higher energy, broad background levels as well as the overall normalization of each data set to accommodate systematic uncertainties in the normalization.

The results from our $R$-matrix analysis are summarized in Table \ref{tab:table3} and in Fig. \ref{fig:summary}. 
Only two resonant features are clearly resolved in all 3 data sets, and we find a good fit to the data with a $1^-$ state describing the lower energy feature near 6.14 MeV. This fit includes a single broad $3^-$ background level at high energies that likely mimics the influence of previously observed $3^-$ states at higher energies \cite{Alm12}. Adding additional broad background states did not improve the quality of the fit in a statistically-significant way.

The resonant feature near $E_x=6.3$~MeV is likely a mixture of the other two previously observed states in the region at 6.29 and 6.35 MeV. 
High-resolution measurements of $^{20}$Ne(p,t)$^{18}$Ne reported intrinsic widths of about 20-45 keV for the states, while they are only separated by about 60 keV \cite{Hah96,Par99}. Considering the experimental resolution, it is not surprising that the 6.29 and 6.35-MeV states are not well resolved, which leads to generally larger uncertainties in the properties of these two levels.

Uncertainties in excitation energies and widths with a single spin subchannel were calculated using Minos \cite{minuitUserGuide} encompassing values satisfying $\chi^2 \leq \chi^2_{\text{min}} + 1.0 $. Total width uncertainties for states that had two spin sub-channels were evaluated by calculating values corresponding to the $\chi^2 = \chi^2_{\text{min}}$ + 2.30 contour and finding the deviation between the width value reported in Table \ref{tab:table3} and the maximum/minimum total widths on the contours.

Only upper limits on widths of one of the two higher-energy states are found in some cases, with the resonance feature being satisfactorily fit with only one of the two levels that are not resolved. In these cases a $\chi^2$ surface was calculated over a grid of values for the resonance energy and partial width of the narrow resonance (in the total spin $S_{\text{tot}}=2$ channel) over a grid of possible values, with an upper limit on the width determined from the largest value within a 68\% confidence interval region ($\chi^2-\chi^2_{\text{min}} <$ 2.30). The resonance parameters obtained from the 3 independent data sets are in agreement, and recommended values are also given in Table \ref{tab:table3}.

\begin{figure}[htp]
	\includegraphics[width=\columnwidth,]{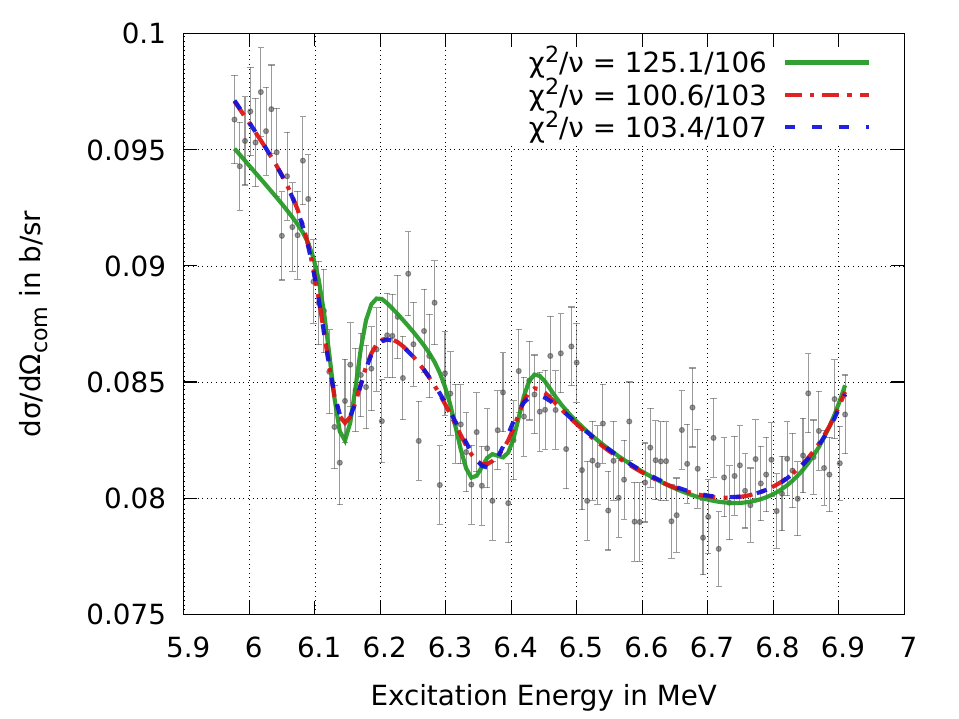}	
	\caption{\label{fig:misc_cases_146} Data from the FSU measurement (same as in Fig. \ref{fig:xsection-146}) but showing 3 different fits each with a $3^-$ assignment for the resonance at 6.14 MeV (Color online). The solid curve (green) describes the alternative fit in Table \ref{tab:table3}, with one background $J^{\pi}=3^-, E_x=8.9$ MeV resonance with freely varying partial widths. The dot-dashed curve (red) describes fit with  resonance parameters of the $4^+$ state adjusted to $E_x=7.065$ MeV, $\Gamma_p=82$ keV, and adds two additional background resonances, with $J^{\pi}=4^+,2^+$, fixed $E_x=8.9$ MeV. The dashed (blue) curve indicates a fit that includes the added $J^{\pi}=2^+$ background resonance used earlier, but fits the region around $E_x=6.35$ MeV with only a $1^-$ state.  }
 \label{fig:misc146-312}
\end{figure}

We also found a different solution that fit the excitation functions with a $3^-$ assignment for the feature at 6.14 MeV. Values for the alternate solution are included in the bottom half of Table \ref{tab:table3} and in Figs. \ref{fig:summary} and \ref{fig:misc_cases_146}. 
The data from Ref. \cite{Hu14} were fit comparably well by either solution.
The fit to the FSU data set using the same assumptions as in the conventional spin-order fit (including a single, broad high-energy $3^-$ background state, and properties for the $4^+$ fixed as described previously) is somewhat worse than the other solution ($\chi^2/\nu=0.99$ for the $1^-$ assignment for the 6.14-MeV state, but $\chi^2/\nu=1.18$ for the $3^-$ assignment for that level). However, the quality of the fit can be improved to be comparable to that with the conventional spin ordering by relaxing particular assumptions, such as adding additional background states or changing parameters of the $4^+$ state. This is illustrated in Fig. \ref{fig:misc146-312} that compares fits to the FSU data set with a $3^-$ assignment for the 6.14-MeV state but using 3 different assumptions.  While the parameter space has to be somewhat expanded to provide a comparable fit with a 3$^-$ assignment for the 6.14-MeV state, this is not strong favoring a particular solution.

\section{Conclusions}
An elastic proton scattering measurement for an unstable beam of $^{17}$F on a thick CH$_2$ target was completed at the RESOLUT facility at FSU.  These measured cross sections, along with previously measured $^{17}$F+p elastic scattering from Ref. \cite{Hu14}, and inelastic scattering measured at ORNL \cite{Bar12} were analyzed in a self-consistent way using the $R$-Matrix formalism to better constrain the properties of resonances at E$_x=6-7$ MeV in $^{18}$Ne that influence the $^{14}$O($\alpha,p$)$^{17}$F reaction.

We find two different sets of resonance parameters that provide a good description of the $^{17}$F+p scattering excitation functions. Weighted-average values for the resonance parameters are provided in Table \ref{tab:table3} and in Fig. \ref{fig:summary}. 
In contrast to the conclusions of Ref. \cite{Hu14}, we find solutions with the 6.14-MeV state having a $J^{\pi}$ of $1^-$ or $3^-$. Test calculations indicate that the differences in our analysis likely arise from the properties used for states outside of the region of interest that change the shape of the resonant interference features. We also find the inelastic scattering data can be described by either solution, as previously reported by Ref. \cite{Bar12}. While mirror symmetry and penetrability arguments favor a $1^-$ assignment for the 6.14-MeV state, with a $3^-$ assignment implying a proton partial width close to the Wigner limit, there is no conclusive evidence supporting either assignment. 

The spin-parity assignments for these states have a significant impact on the $^{14}$O($\alpha,p$)$^{17}$F reaction rate due to the difference in orbital angular momentum and penetrability through the Coulomb barrier. Assuming the same reduced alpha widths ($\Gamma_{\alpha}$) for these states, the reaction rate for the solution with the 6.14-MeV state having a $3^-$ assignment is more than a factor of 6 smaller at $T=0.5$~GK than the rate implied by the other solution. However, near peak X-ray burst temperatures (2 GK), the rate with a $3^-$ assignment for the 6.14-MeV state is about a factor of 2 larger. Additional evidence is needed to clearly establish the mirror assignments and reduce the uncertainty in the $^{14}$O($\alpha,p$)$^{17}$F reaction rate.

\section{Acknowledgements}
The authors thank Richard deBoer, Daniel Odell, Thomas King, and Amber Lauer-Coles for clarifying discussions.  This work was partially supported by the U.S. Department of Energy, Office of Science under Grant No. DE-FG02-96ER40989 and by the National Science Foundation under Grant Nos. PHY-2012522 and PHY-2310059. Portions of this research were conducted with high performance computing resources provided by Louisiana State University \cite{lsuhpc}.

\bibliographystyle{apsrev4-2}
\bibliography{apssamp2-1}
\end{document}